\documentclass[twocolumn,showpacs,superscriptaddress,groupedaddress,letterpaper]{revtex4}
\usepackage{amsfonts}
\usepackage{graphicx}
\usepackage{dcolumn}
\usepackage{bm}
\usepackage{amssymb}
\usepackage{amsmath}
\usepackage{slashed}
\usepackage[pdfborder={0 0 0},colorlinks=true,linkcolor=blue,citecolor=blue,filecolor=blue,urlcolor=blue,breaklinks=true]{hyperref}

\begin{document}

\title{Comment on ``Chern-Simons theory and atypical Hall conductivity \\ in
the Varma phase''}
\author{Angel E. Obispo}
\email{angel.obispo@ufma.br, ae.obispo@gmail.com}
\affiliation{UFMA Universidade Federal do Maranh\~{a}o - Campus do Bacanga - CCET,
65080-805, S\~{a}o Lu\'{\i}s-MA, Brazil}
\author{Francisco A. Cruz Neto}
\affiliation{UFMA Universidade Federal do Maranh\~{a}o - Campus do Bacanga - CCET,
65080-805, S\~{a}o Lu\'{\i}s-MA, Brazil}
\author{Andr\'{e}s G. Jir\'{o}n Vicente}
\affiliation{UNAC Universidad Nacional del Callao - Campus Central - FCNM. 07001,
Bellavista - Callao, Per\'{u}}
\affiliation{UFMA Universidade Federal do Maranh\~{a}o - Campus do Bacanga - CCET,
65080-805, S\~{a}o Lu\'{\i}s-MA, Brazil}
\author{L. B. Castro}
\email{luis.castro@ufma.br}
\affiliation{UFMA Universidade Federal do Maranh\~{a}o - Campus do Bacanga - CCET,
65080-805, S\~{a}o Lu\'{\i}s-MA, Brazil}
\affiliation{UNESP Universidade Estadual Paulista - Campus de Guaratinguet\'{a} - DFQ,
12516-410, Guaratinguet\'{a}-SP, Brazil}

\begin{abstract}
In a recent paper published in this Journal [Phys. Rev. B \textbf{97},
075135 (2018)], Menezes \emph{et al. } analyze the topological behavior of a
effective bosonic model defined on the Lieb lattice in presence of an
electromagnetic field. In this context, the authors claim to have found an
atypical quantum Hall effect for the quasiparticles. However, some
inconsistencies related to the treatment of the propagator jeopardizes the
main result in this system.
\end{abstract}

\maketitle




In an interesting Letter, Menezes \emph{et al. }\cite{PRB97:075135:2018}\emph{\ }%
analized the topological response of a effective bosonic theory defined on
the Lieb lattice which is minimally coupled to an external $U(1)~$gauge
field in $2+1~$dimensions. To this purpose, the authors consider a
tight-binding hamiltonian with three different species of (pseudo-) gapped
fermions (see equations (1)-(3) in Ref.\cite{PRB97:075135:2018}), similar to the one
proposed in Ref.\cite{PRB85:155106:2012}. Such pseudo-gap behavior arises from the
so-called Varma phase \cite{PRL83:3538:1999} which break time-reversal
symmetry spontaneously (preserving the translational symmetry of the
lattice) and whose realization would be possible in the copper-oxygen planes
of high temperature cuprate superconductors \cite{PRL83:3538:1999,PRB73:155113:2006}.

As a first result, in the Section II in Ref.\cite{PRB97:075135:2018}, the authors
showed that the dynamics of the charge carriers on that Lieb lattice in
low-energy regime present a relativistic-like behavior, correctly described
by Duffin-Kemmer-Petiau-like hamiltonian (DKP). Disregarding irrelevant
constants, this effective hamiltonian is expressed in a simplified version
as (see equation (4) in Ref.\cite{PRB97:075135:2018})%
\begin{equation}
H_{DKP}\Psi =E\Psi ,  \label{1.0}
\end{equation}%
\begin{equation}
H_{DKP}=\left[ \beta ^{0},\beta ^{1}\right] k^{1}+\left[ \beta ^{0},\beta
^{2}\right] k^{2}+m\beta ^{0},  \label{1}
\end{equation}%
where $\Psi \ $is the three-component spinor (see figure (1) in Ref.\cite{PRB97:075135:2018}),%
\begin{equation}
\Psi (\mathbf{k})=\left(
\begin{array}{c}
b(\mathbf{k}) \\
a(\mathbf{k}) \\
c(\mathbf{k})%
\end{array}%
\right) ,  \label{1.1}
\end{equation}%
$\mathbf{k=(}k^{1},k^{2}\mathbf{)}~$is the momentum or wave vector, and $%
\beta ^{i}~$are two $3\times 3~$anti-hermitian matrices which, together with
another $3\times 3~$ hermitian matriz $\beta ^{0}$, satisfy the so-called DKP
algebra%
\begin{equation}
\beta ^{\mu }\beta ^{\nu }\beta ^{\sigma }+\beta ^{\sigma }\beta ^{\nu
}\beta ^{\mu }=\beta ^{\mu }\eta ^{\nu \sigma }+\beta ^{\sigma }\eta ^{\nu
\mu },  \label{2}
\end{equation}%
with the metric tensor $\eta ^{\mu \nu }=$diag$(1,-1,-1)$. This is the basis
on which \cite{PRB97:075135:2018} develops.

In simple terms, the DKP equation is a first-order wave equation that
defines spin 0 (scalar sector) and spin 1 (vectorial sector) fields\ and
particles with a rich algebraic structure not capable of being expressed in
the traditional Klein-Gordon (KG) and Proca theories \cite{PRA90:022101:2014}. Related
to this, the authors in Ref.\cite{PRB97:075135:2018} state that the charge carriers on
the Lieb lattice are described by \textit{relativistic pseudospin-0
quasiparticles in two spatial dimensions}, i.e, these would exist in the
scalar sector of the theory. We rebut this statement via the following
argument. In $3+1~$spacetime dimensions, the algebra (\ref{2}) generates a
set of 126 independent matrices whose irreducible representations are a
trivial representation, a five-dimensional representation for the scalar
sector, and a ten-dimensional representation for the vectorial sector \cite%
{PRA90:022101:2014,CORSON1953}. Whatever the sector, it is clear that the DKP spinor will
have an excess of components. In this case, the theory needs to be
complemented by a constraint equation that
allows to eliminate the redundant components, which is given by%
\begin{equation}
\beta ^{i}\beta ^{0}\beta ^{0}k_{i}\Psi =m\left( 1-\beta ^{0}\beta
^{0}\right) \Psi ,~~~i=1,2,3.  \label{3}
\end{equation}

With this constraint equation we can express the three (four) components of
the spinor by the other two (six) components and their space derivatives in
the scalar (vector) sector - for more detail see Ref.\cite{PRA90:022101:2014}. Thereby,
we can exclude the redundant components and reexpress our system of
equations to another that depends only on physical components (1 for the
scalar sector and 4 for the vectorial sector) of the DKP theory. If we
performed this same analysis in \thinspace $2+1~$dimensions, we will see
that the algebra (\ref{2}) now generates a set of 35 independent matrices
whose irreducible representations are a trivial representation, a
four-dimensional representation, and two different three-dimensional
representations. Note that only the three-dimensional representations adapt
to the structure of the Hamiltonian (\ref{1}), reducing $\Psi ~$to a
three-component spinor, similar to (\ref{1.1}). As the Lieb lattice has three bands at low-energy and we
have three components to the spinor in (\ref{1.1}), these components come from
the three-component pseudo-spin 1 quasi-particles. The constraint
equation (\ref{3}) (modified to two space dimensions, $i=1,2$) eliminates the
redundant pseudo-spin 0 quasi-particles. Therefore, (\ref{1.1}) describes pseudospin-1
quasiparticles. It is important to highlight that constraint equation also allows one to demonstrate the
equivalence between the hamiltonian form (\ref{1}) and the DKP equation of
motion, which is not a trivial matter \cite{PRA90:022101:2014,PLA268:165:2000}. In fact, a DKP
quantum field theory is possible only\ if this equivalence is established,
as can be verified in Refs.\cite{AKHIEZER1965,VALVERDE2000,IJMPA26:2487:2011}.

The section III in Ref.\cite{PRB97:075135:2018} represents the crucial point of this
Comment. In that section, the authors analyzed the topological response
generated by one-loop radiative corrections to the two-point function of the
gauge and DKP fields in $2+1$ dimensions. As it is known from usual QED$%
_{2+1}$, such topological term - called Chern-Simons term - comes from the
first-order in external momentum contribution of the vacuum polarization
diagram \cite{PRB89:165405:2014}. In Physics of Condensed Matter (perhaps its most
notable application), this emergent Chern-Simons theory naturally leads to
the transverse conductivity observed from the Hall effect. In this context,
the authors claim to have found\ \textit{an atypical quantum Hall effect for
the DKP quasiparticles }(eq. (15) in Ref.\cite{PRB97:075135:2018}), given by%
\begin{equation}
\sigma ^{xy}=\text{sign}(m)\frac{q^{2}}{4h}.  \label{rp}
\end{equation}%
\

The above expression is their main result and represents a truly atypical
result (one should expect to obtain an integer quantum hall effect), which
shows for the first time in both condensed matter and higher-energy physics
literature the derivation of an Abelian Chern-Simons theory from a non-Dirac
system (namely DKP system). Unfortunately, we found some inconsistencies
related to the treatment of the DKP propagator, in fact, the expression (13)
in Ref.\cite{PRB97:075135:2018} is incorrect \cite{AKHIEZER1965,VALVERDE2000,IJMPA26:2487:2011}, therefore,
the final result (\ref{rp}) is invalid. Below we justify our statement.

\begin{figure}[t]
\begin{center}
\includegraphics[width=6cm, angle=0]{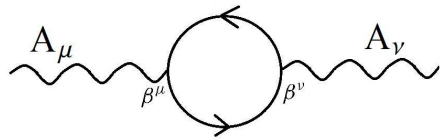}
\end{center}
\par
\vspace*{-0.1cm}
\caption{Vacuum polarization diagram}
\label{figura1}
\end{figure}

We start considering that the interaction of the electromagnetic field with
the long wavelength (low-frequency) excitations of charge carriers on the
Lieb lattice can be described by relativistict quantum electrodynamics for
integer spin particles \cite{AKHIEZER1965,VALVERDE2000,IJMPA26:2487:2011}. The action (8) in Ref.%
\cite{PRB97:075135:2018} is built upon the motion equation%
\begin{equation}
\left( i\hbar \slashed{\partial}-q\slashed{A}-m\right) \Psi =0,  \label{EM}
\end{equation}%
where $m~$is the mass-gap parameter, $q~$is coupling parameter, $A_{\mu }~$%
is the vector gauge potential,
\begin{equation*}
\slashed{\partial}=\beta ^{\mu }\frac{\partial }{\partial x^{\mu }},~\ \ %
\slashed{A}=\beta ^{\mu }A_{\mu }.\
\end{equation*}%
The resulting polarization tensor in the momentum representation is given by
\begin{equation}
i\Pi ^{\mu \nu }\left( p\right) =+\frac{q^{2}}{\hbar }\int \frac{d^{3}k}{%
(2\pi )^{3}}\text{Tr}\left[ \beta ^{\mu }G_{\Psi }(k-p)\beta ^{\nu }G_{\Psi
}(k)\right] ,  \label{4}
\end{equation}%
where
\begin{equation}
G_{\Psi }(k)=i\frac{1}{\beta ^{\mu }k_{\mu }-m},  \label{5}
\end{equation}%
is the DKP free (Feynman) propagator. The tensor polarization (\ref{4}) is
equivalent to equation (12) in Ref.\cite{PRB97:075135:2018}, by changing $\mu
\leftrightarrow $ $\nu $. The vacuum polarization diagram we have considered
is the one shown in Fig.(\ref{figura1}), which allowed us built $i\Pi ^{\mu
\nu }\left( p\right) ~$following the same Feynman rules of the usual QED$%
_{2+1}$, except for the plus sign ($+$) in front, which is a reminiscent of the bosonic nature of the
DKP theory \cite{AKHIEZER1965}. The standard procedure to calculate $i\Pi ^{\mu \nu
}\left( p\right) $ says that we must first evaluate the trace of $\beta ~$%
matrices, which implies that $G_{\Psi }(k)~$in (\ref{5}) must be rewritten
in such a way that these matrices appear in the numerator. Nevertheless,
this process in DKP theory is more complicated (as compared with Dirac
theory) due mainly to its algebra and because the $\beta ~$matrices are
singulars (det$[\beta ]=0$). As $\beta ^{-1}~$does not exist, it implies
that some common identities are not valid anymore, for instance, $\left(
\beta ^{\mu }p_{\mu }\right) \left( \beta ^{\nu }p_{\nu }\right) ^{-1}=%
\mathbb{I}$.\ This did not take into account by the authors in Ref.\cite%
{PRB97:075135:2018}, and as result, their propagator was incorrectly constructed (see
equation (13) in Ref.\cite{PRB97:075135:2018}). The correct form for the DKP
propagator is also performed in Refs.\cite{IJMPA17:205:2002,JPCS706:052002:2016,PRD101:065003:2020}, and is
expressed as follows \
\begin{equation}
G_{\Psi }(k)=i\frac{1}{\beta ^{\mu }k_{\mu }-m}=\frac{i}{m}\left[ \frac{%
\slashed{k}(\slashed{k}+m)}{k^{2}-m^{2}}-1\right] .  \label{6}
\end{equation}

It is straightforward to see that the propagator in (\ref{6}) is strictly
defined for massive particles, as required for the DKP theory. In fact, the
equation (\ref{1}) with $m=0~$represent a different relativistic equation,
the so-called Harish-Chandra equation \cite{PRSLA186:502:1946}, whose analysis we will
leave aside here. So, we go to focus on the propagator in (\ref{6}),
alternatively rewritten as%
\begin{eqnarray}
G_{\Psi }(k) &=&G_{1}(k)+G_{2}(k)  \notag \\
&=&i\frac{(\slashed{k}+m)}{k^{2}-m^{2}}+\frac{i}{m}\frac{(\slashed{k}%
^{2}-k^{2})}{k^{2}-m^{2}}.  \label{7}
\end{eqnarray}

Note that the first term on the right $G_{1}(k)~$coincides exactly with the
propagator proposed by Menezes \emph{et al. }(see equation (13) in Ref.\cite%
{PRB97:075135:2018}), which is used to determine the value of the Hall conductivity by
considering the contribution from antisymmetric part of the polarization
tensor $i\Pi ^{\mu \nu }\sim \int d^{3}k$Tr$\left[ \beta ^{\mu }G_{1}\beta
^{\nu }G_{1}\right] $ and the second term $G_{2}(k)$ absent. The point of
this Comment is to demonstrate that, when $G_{2}(k)$ is inserted into (\ref%
{4}), the \newline term $\sim \int d^{3}k$Tr$\left[ \beta ^{\mu }G_{1}\beta ^{\nu
}G_{2}+\beta ^{\mu }G_{2}\beta ^{\nu }G_{1}+\beta ^{\mu }G_{2}\beta ^{\nu
}G_{2}\right]$ contributes a non-negligible value to the value of the Hall
conductivity found in Ref.\cite{PRB97:075135:2018}.

For this purpose, we rewrite the polarization tensor according to
decomposition (\ref{7})%
\begin{equation}
i\Pi ^{\mu \nu }\left( p\right) =i\Pi _{(1,1)}^{\mu \nu }+i\Pi _{(1,2)}^{\mu
\nu }+i\Pi _{(2,1)}^{\mu \nu }+i\Pi _{(2,2)}^{\mu \nu },  \label{8}
\end{equation}%
where $i\Pi _{(i,j)}^{\mu \nu }$ are functions of momentum $p$, conveniently defined as%
\begin{eqnarray*}
i\Pi _{(1,1)}^{\mu \nu } &=&\frac{q^{2}}{\hslash }\int \frac{d^{3}k}{(2\pi
)^{3}}\text{Tr}\left[ \beta ^{\mu }G_{1}(k-p)\beta ^{\nu }G_{1}(k)\right]
,~~~~~ \\[0.08in]
i\Pi _{(1,2)}^{\mu \nu } &=&\frac{q^{2}}{\hslash }\int \frac{d^{3}k}{(2\pi
)^{3}}\text{Tr}\left[ \beta ^{\mu }G_{1}(k-p)\beta ^{\nu }G_{2}(k)\right]
,~~~~~ \\[0.08in]
i\Pi _{(2,1)}^{\mu \nu } &=&\frac{q^{2}}{\hslash }\int \frac{d^{3}k}{(2\pi
)^{3}}\text{Tr}\left[ \beta ^{\mu }G_{2}(k-p)\beta ^{\nu }G_{1}(k)\right]
,~~~~~ \\[0.08in]
i\Pi _{(2,2)}^{\mu \nu } &=&\frac{q^{2}}{\hslash }\int \frac{d^{3}k}{(2\pi
)^{3}}\text{Tr}\left[ \beta ^{\mu }G_{2}(k-p)\beta ^{\nu }G_{2}(k)\right]
.~~~~~
\end{eqnarray*}
As previously mentioned, the result obtained in Ref.\cite{PRB97:075135:2018} can be
reproduced by computing simply the antisymmetric part of $i\Pi _{(1,1)}^{\mu
\nu }$, so, in that sense, it is convenient to divide the Eq. (\ref{8}) into
two contributions: $i\Pi _{(1,1)}^{\mu \nu }~$and $i\Pi _{(1,2)}^{\mu \nu
}+i\Pi _{(2,1)}^{\mu \nu }+i\Pi _{(2,2)}^{\mu \nu }$. As we are only
interested in the topological part of each contribution, we focus our
attention on the terms $\sim \epsilon ^{\mu \nu \alpha }p_{\alpha }$,$~$%
which come from combining an odd number of $\beta $ matrices,%
\begin{eqnarray}
\text{Tr}\left[ \beta ^{\rho }\beta ^{\sigma }\beta ^{\theta }\right]
&=&i\epsilon ^{\rho \sigma \theta },  \label{9.1} \\[0.2cm]
2\text{Tr}\left[ \beta ^{\rho }\beta ^{\alpha }\beta ^{\sigma }\beta
^{\omega }\beta ^{\theta }\right] &=&ig^{\rho \alpha }\epsilon ^{\sigma
\omega \theta }+ig^{\alpha \sigma }\epsilon ^{\omega \theta \rho }  \notag \\%
[0.18cm]
&&+ig^{\sigma \omega }\epsilon ^{\theta \rho \alpha }+ig^{\omega \theta
}\epsilon ^{\rho \alpha \sigma }  \notag \\[0.2cm]
&&+ig^{\rho \omega }\epsilon ^{\alpha \sigma \theta }+ig^{\alpha \theta
}\epsilon ^{\rho \sigma \omega }.  \label{9.2}
\end{eqnarray}
Thus, following the standard methods for QED calculations \cite{AKHIEZER1965}, we will
computed the antisymmetric part of each contribution separately.\medskip
\begin{itemize}
\item \textbf{Computing $i\Pi _{AS(1,1)}^{\mu \nu }$:}
\end{itemize}
To determinate the contribution from this first term (which is used in Ref.\cite{PRB97:075135:2018} to get its main result), we start applying the trace property (\ref{9.1}). After a few calculations we obtain
\begin{equation*}
i\Pi _{AS(1,1)}^{\mu \nu }=-\frac{imq^{2}}{\hslash }\int_{0}^{1}dx\int \frac{%
d^{3}k}{(2\pi )^{3}}\frac{p_{\alpha }}{\left[ k^{2}-\Delta ^{2}\right] ^{2}}%
\epsilon ^{\mu \nu \alpha },
\end{equation*}%
where we have used the Feynman parametrization procedure
\begin{equation}
\frac{1}{\left[ \left( k-p\right) ^{2}-m^{2}\right] \left[ k^{2}-m^{2}\right]
}=\int_{0}^{1}dx\frac{1}{\left[ k^{2}-\Delta ^{2}\right] ^{2}},  \label{10}
\end{equation}%
together with the change $k\rightarrow k+xp~$and $\Delta =m^{2}-p^{2}x(1-x)$%
. These integrals are the massive one-loop Feynman integrals, widely studied
in QED, and whose result we will use directly. Thereby, we get
\begin{eqnarray}
i\Pi _{AS(1,1)}^{\mu \nu } &=&\frac{m}{\left\vert m\right\vert }p_{\alpha
}\epsilon ^{\mu \nu \alpha }\frac{q^{2}}{4h }\int_{0}^{1}dx\frac{1}{\sqrt{%
1-x\left( 1-x\right) p^{2}/m^{2}}},  \notag \\[0.2cm]
&=&\text{sgn}(m)\frac{1}{4}\frac{q^{2}}{h}p_{\alpha }\epsilon ^{\mu \nu
\alpha }, \label{11.1}
\end{eqnarray}
where in the last line we have considered the Chern-Simons regime ($m\gg p$%
). This is the result obtained by Menezes in their manuscript, as expected.\medskip
\begin{itemize}
\item \textbf{Computing $i\Pi _{AS(1,2)}^{\mu \nu }+i\Pi _{AS(2,1)}^{\mu \nu
}+i\Pi _{AS(2,2)}^{\mu \nu }$:}
\end{itemize}
In this case, both (\ref{9.1}) and (\ref{9.2}) are required. A quick
inspection allows us to demonstrate that $i\Pi _{AS(2,2)}^{\mu \nu }=0$,
i.e, this term does not have a antisymmetric part. By other hand, the sum of
the cross terms provides%
\begin{eqnarray*}
i\Pi _{AS(1,2)}^{\mu \nu }+i\Pi _{AS(2,1)}^{\mu \nu }&=&-\frac{iq^{2}}{2m\hslash }\int_{0}^{1}dx\int \frac{d^{3}k}{(2\pi )^{3}}\times  \\
&&\frac{1}{\left[ \left( k-px\right) ^{2}-\Delta ^{2}\right]^{2}}\times\text{Tr}\left[p\right],
\end{eqnarray*}
where we use the parametrization (\ref{10}), with $\Delta =m^{2}-p^{2}x(1-x)~
$and%
\begin{eqnarray*}
\text{Tr}\left[p\right]  &=&2\epsilon ^{\nu \alpha \omega }p_{\alpha
}k^{\mu }k_{\omega }-2\epsilon ^{\mu \alpha \omega }p_{\alpha }k^{\nu
}k_{\omega } \\[0.2cm]
&&+2\epsilon ^{\mu \nu \alpha }k_{\alpha }\left( pk\right) -\epsilon ^{\nu
\alpha \omega }p_{\alpha }p^{\mu }k_{\omega } \\[0.2cm]
&&+\epsilon ^{\mu \alpha \omega }p_{\alpha }p^{\nu }k_{\omega }-\epsilon
^{\mu \nu \alpha }p_{\alpha }\left( pk\right) -\epsilon ^{\mu \nu \alpha
}k_{\alpha }p^{2}.
\end{eqnarray*}
At this point, a regularization scheme is required. We use the Pauli-Villars
regularization, which allow us to make the change $k\rightarrow k+px$, to
then exclude the linear terms in $k$ associated to odd-integrals, and
replace $k^{\mu }k^{\nu }$ by $g^{\mu \nu }k^{2}/3$ in numerator. Thereby we
obtain:
\begin{eqnarray*}
i\Pi _{\text{AS}(1,2)}^{\mu \nu }+i\Pi _{\text{AS}(2,1)}^{\mu \nu }&=&\text{sgn}(m)\frac{3}{4}\frac{q^{2}}{h}\epsilon ^{\mu \nu \alpha }p_{\alpha}\times\\
&&\int_{0}^{1}dx\frac{3-2x\left( 1-x\right) \left( p^{2}/m^{2}\right) }{3\sqrt{1-x\left( 1-x\right) \left( p^{2}/m^{2}\right) }}.
\end{eqnarray*}%
In the Chern-Simons regime ($m\gg p$)
\begin{equation}
i\Pi _{\text{AS}(1,2)}^{\mu \nu }\left( p\right) +i\Pi _{\text{AS}%
(2,1)}^{\mu \nu }\left( p\right) =\text{sgn}(m)\frac{3}{4}\frac{q^{2}}{h}%
\epsilon ^{\mu \nu \alpha }p_{\alpha }.  \label{12}
\end{equation}
The above expression represent the main result of this Comment. Note that it is three times larger than the one found in (\ref{11.1}), which implies a important modification to the result found by Menezes \emph{et al.} \cite{PRB97:075135:2018}.
Thus, if we combine the eqs. (\ref{11.1}) and (\ref{12}) to find the full expression of %
the dynamically generated Chern-Simons term, we get%
\begin{eqnarray}
i\Pi _{AS}^{\mu \nu }\left( p\right) &=&i\Pi _{AS(1,1)}^{\mu \nu }+i\Pi
_{AS(1,2)}^{\mu \nu }+i\Pi _{AS(2,1)}^{\mu \nu },\notag\\[0.17cm]
&=&\text{sign}(m)\frac{q^{2}}{h}\epsilon ^{\mu \nu \theta }p_{\theta }.\label{Tas}
\end{eqnarray}%

The Hall conductivity can be obtained via the Kubo's formula%
\begin{equation}
\sigma^{xy}=\lim_{\mathbf{p}\rightarrow 0,~p_{0}\rightarrow 0}\frac{i\Pi
_{AS}^{xy}}{p_{0}}=\text{sign}(m)\frac{q^{2}}{h},  \label{Hall}
\end{equation}%
which is the result expected according to the literature \cite{PRL83:3538:1999,PRB87:125428:2013,PRA83:063601:2011}. Therefore, there is no such atypical Hall conductivity, as
reported by the authors.

We conclude by emphasizing that the results and conclusions presented is
this Comment do not alter the others results shown in \cite{PRB97:075135:2018},
concerning the obtaining of Landau levels in DKP theory and to the extension
of Jackiw-Rebbi approach for the DKP quasiparticles.

\section*{ACKNOWLEDGMENTS}

Angel E. Obispo thanks to CNPq (grant 312838/2016-6) and Secti/FAPEMA (grant
FAPEMA DCR-02853/16), for financial support. L. B. Castro also thanks to
CNPq, Brazil, Grant No. 307932/2017-6 (PQ) and No. 422755/2018-4
(UNIVERSAL), S\~{a}o Paulo Research Foundation (FAPESP), Grant No.
2018/20577-4, FAPEMA, Brazil, Grant No. UNIVERSAL-01220/18 and CAPES, Brazil. The authors would like to
thank R. Casana for useful discussions.

\bibliographystyle{spphys}       

\end{document}